# Evolutionary Dynamics of Scientific Collaboration Networks: Multi-Levels and Cross-time Analysis

*Alireza Abbasi [1], Liaquat Hossain [1], Shahadat Uddin [1], Kim J.R. Rasmussen [2]*

[1] Centre for Complex Systems Research, Faculty of Engineering and IT, University of Sydney, Australia
[2] School of Civil Engineering, Faculty of Engineering and IT, University of Sydney, Australia

**Abstract**

Several studies exist which use scientific literature for comparing scientific activities (e.g., productivity, and collaboration). In this study, using co-authorship data over the last 40 years, we present the evolutionary dynamics of multi level (i.e., individual, institutional and national) collaboration networks for exploring the emergence of collaborations in the research field of "steel structures". The collaboration network of scientists in the field has been analyzed using author affiliations extracted from Scopus between 1970 and 2009. We have studied collaboration distribution networks at the micro-, meso- and macro-levels for the 40 years. We compared and analyzed a number of properties of these networks (i.e., density, centrality measures, the giant component and clustering coefficient) for presenting a longitudinal analysis and statistical validation of the evolutionary dynamics of "steel structures" collaboration networks. At all levels, the scientific collaborations network structures were central considering the *closeness* centralization while betweenness and degree centralization were much lower. In general networks density, connectedness, centralization and clustering coefficient were highest in marco-level and decreasing as the network size grow to the lowest in micro-level. We also find that the average distance between countries about two and institutes five and for authors eight meaning that only about eight steps are necessary to get from one randomly chosen author to another.

## 1. Introduction

In recent years, there has been a sharp increase in the number of collaborations between scholars. An explanation for the rapid growth of international scientific collaboration has been provided by Luukkonen and colleagues (1992; 1993) as well as Wagner and Leydesdorff (2005). By jointly publishing a paper, researchers show their knowledge-sharing activities, which are essential for knowledge creation. "The rising awareness of collaborativeness in science has led to a sharpened focus on the collaboration issue" (Melin 2000). Scientific collaboration has even been called a "springboard for economic prosperity and sustainable development" (US Office of Science & Technology Policy 2000). As most scientific output is a result of group work and most research projects are too large for an individual researcher to perform, scientific cooperation between individuals across national borders is often required to develop a holistic approach to the phenomenon under investigation (Leclerc and Gagné 1994).

Since scientific collaborations are defined as "interactions taking place within a social context among two or more scientists that facilitate the sharing of meaning and completion of tasks with respect to a mutually shared, super-ordinated goal" (Sonnenwald 2007), those collaborations frequently emerge from, and are perpetuated through, social networks. Since social networks may span disciplinary, institutional, and national boundaries, it can influence collaboration in multiple ways (Sonnenwald 2007). Social network analysis has produced many results concerning social influence, social groupings, inequality, disease propagation, communication of information, and indeed almost every topic that has interested 20th century sociology (Newman 2001).

Social networks operate at many levels, from families up to the level of nations. They play a critical role in determining the way problems are solved, institutions are run, markets evolve, and the degree to which individuals succeed in achieving their goals (Abbasi and Altmann 2011). Social networks have been analyzed to identify areas of strengths and weaknesses within and among research institutions, businesses, and nations as well as to direct scientific development and funding policies (Owen-Smith, Riccaboni et al. 2002; Sonnenwald 2007).

A social network can be conceptualized as a set of individuals or groups, each of which has connections of some kind to some or all of the others. In the language of social network analysis, people or groups are called ''actors'' or "nodes" and connections are referred to as ''ties'' or "links". Both actors and ties can be defined in different ways depending on the questions of interest. An actor might be a single person, a team, or a company. A tie might be a friendship between two people, collaboration or common member between two teams, or a business relationship between companies (Newman 2001). In scientific collaborations' network actors (nodes) are authors and ties (links) are co-authorship relations among them. A tie exists between each two actors (scholars) if they have at least one co-authored paper.

Constructing collaboration (co-authorship) networks of scholars is being widely used to analyse the network structure and actors position and attributes (Melin and Persson 1996; Katz and Martin 1997; Newman 2001; Newman 2001; Barabási, Jeong et al. 2002; Grossman 2002; Moody 2004; Newman 2004; Acedo, Barroso et al. 2006; Sonnenwald 2007; Suresh, Raghupathy et al. 2007; Jiang 2008; Abbasi, Altmann et al. 2010; Abbasi, Altmann et al. in press) for different domains but to our knowledge there is no such study of "steel structures" research collaboration networks. In this study, based on the affiliation data of scholars as obtained from publications, we applied several network levels (country, institute and author) of analysis to show the current state of the collaboration network. Furthermore, evolution of the network since the initiation publications (the first node of the network in our dataset) and investigating fast growing countries in this field have been explored.

In addition to measuring the number of actor collaborators and the frequency of the collaborations with each actor, as widely used in similar literature, we also measure the following other quantities of collaboration networks:

**Network Density:**

Density describes the general level of linkage among the nodes in a network (Scott 1991). The more nodes are connected to one another, the denser the network is. Density describes the general level of cohesion in a network (Scott 1991). More specifically, density of a network is the proportion of exiting links to the maximum possible number of distinct links.

**Network Centralization:**

Another method used to understand networks and their participants is to evaluate the location of actors in the network. Measuring the network location is about determining the centrality of an actor. These measures help determine the importance of a node in the network. "Centralization describes the extent to which this cohesion is organized around particular focal nodes" (Scott 1991). Bavelas (1950) was the pioneer who initially investigates formal properties of centrality and proposed several centrality concepts. Later, Freeman (1979) found that centrality is an important structural factor influencing leadership, satisfaction, and efficiency.

To examine if a whole network has a centralized structure, centralization can be used. It refers to 'compactness' of a network. A network's centralization indicates how tightly the network is organized around its most central nodes. The general view is finding differences between most central nodes' centrality scores and others'. Then, centralization is calculated as the ratio of the sum of these differences to the maximum possible sum of differences. Therefore, to calculate a network centralization, the first step is to find all nodes measures and then find the whole network centralities measures. The important node centrality measures are:

- *Degree Centrality*

The degree centrality is simply the number of other nodes connected directly to a node. Necessarily, a central node is not physically in the centre of the network as degree of a node is calculated in terms of the number of its adjacent nodes.

- *Closeness Centrality*

Freeman (1979; 1980) proposed closeness in terms of the distance among various nodes. Extending the same concept Sabidussi (1966) used in his work the 'sum distance', the sum of the 'geodesic' distances (the shortest path between any particular pair of nodes in a network) to all other nodes in the network. A node is globally central if it lies at the shortest distance from many other nodes. However, simply calculating the sum of distances of a node to other nodes will produce a measure of 'farness' and so we need to use the inverse of the 'farness' measure as a measure of closeness. In unconnected networks, every node is at an infinite distance from at least one node, and closeness centrality of all nodes would be 0. Thus, in order to solve this problem to consider all nodes, Freeman proposed to calculate closeness of a node as the "*sum of reciprocal distances*" of a particular node to any other nodes.

- *Betweenness Centrality*

Freeman (1979) also proposed a concept of centrality which measures the number of times a particular node lies 'between' the various other nodes in the network. Betweenness centrality is defined more precisely as the number of shortest paths (between all pairs of nodes) that pass through a given node (Borgatti 1995).

**The Giant Component:**

In small networks (few nodes and connections), all individuals belong to a small group of collaborations or communications. As the total number of connections increases, however, there comes a point at which a "giant component" forms, i.e., "a large group of individuals who are all connected to one another by paths of intermediate acquaintances" (Newman 2001). It is important to realize that a collaboration network is usually fragmented in many clusters (components). One of the reasons for this is that in every field there are scientists that do not collaborate at all, that is they are single authors of all papers on which their names appear. In most research fields, apart from a very small fraction of authors who do not collaborate, all authors belong to a single "giant component" (cluster) from the very early stages of the field (Barabási, Jeong et al. 2002).

**Clustering Coefficients:**

Networks are mostly clustered which means they possess local communities in which a higher than average number of people know one another. One way to check the existence of such clustering in network data is to measure the fraction of ''transitive triples'' (also called clustering coefficients) in a network (Newman 2001). The clustering coefficients of a network is the fraction of ordered triples of nodes A, B, C in which edges AB and BC are present that have edge AC present. In other words, it is the probability that two neighbors of a vertex are adjacent to each other. Clustering coefficient is an important property of networks which is "the probability that two of a scientist's collaborators have themselves collaborated" (Barabási, Jeong et al. 2002; Grossman 2002).

Watts and Strogatz (1998) defined clustering coefficient as follows. Consider node $i$ that has links to $k_i$ other nodes in the network. If these $k_i$ nodes form a fully connected sub-network (clique), there are $k_i(k_i-1)/2$ links between them, but in reality we find much never. Let us denote by $N_i$ the number of links that connect the selected $k_i$ nodes to each other. The clustering coefficient for node $i$ is then $C_i = 2N_i / k_i(k_i-1)$. In simple terms, the clustering coefficient of a node in the co-authorship network tells us how much a node's collaborators are willing to collaborate with each other, and it represents the probability that two of its collaborators wrote a paper together (Barabási, Jeong et al. 2002). Thus, authors' low clustering coefficients value means less probability of collaboration among the non-connected co-authors of the author. This gives power to the author by brokering new collaborations between their co-authors. The clustering coefficient for the whole network is obtained by averaging $C_i$ over all nodes in the system.

## 2. Data Sources

Scopus (www.scopus.com) is one of the main sources of bibliometric data. To construct the database for this study, publications were extracted using the string "steel structure" in the titles or keywords or abstracts in the top 15 specified journals of the field (shortlisted by one of authors as an expert of the field) and restricting the search to publications in English. The 15 specified journals are: *Journal of Constructional Steel Research; Journal of Structural Engineering; Engineering Structures; Thin-Walled Structures; Computers and Structures; Journal of Engineering Mechanics; Earthquake Engineering and Structural Dynamics; Fire Safety Journal; Canadian Journal of Civil Engineering; International Journal of Impact Engineering; Engineering Fracture Mechanics; Fatigue and Fracture of Engineering Materials and Structures; Structural Engineer; Advances in Structural Engineering;* and *Steel and Composite Structures.*

After extracting the publications' meta-data from Scopus and importing the information (i.e., title, publication date, author names, affiliations, publisher, number of citations, etc), we used an application program, AcaSoNet (Abbasi and Altmann 2010), for extracting relationships (e.g., co-authorships) between researchers, and stored the data in tables in a local relational database. Four different types of information were extracted from each publication meta-data: Publications information (i.e., title, publication date, journal name, etc); authors' names; affiliations of authors (including country, institute and department name, etc); and keywords.

As we are interested in different macro-, meso- and micro-levels of analysis (i.e., country, institute, author), affiliation data is so important for our research. We found affiliation information, in our original extracted data, especially for older publications messy (the order and even different written name for the country, institute and department, etc) and also for some publications some fields (e.g., country) were missing. So, in our second step we undertook manual checks (using Google) to fill the missing fields using other existing fields (e.g., we used institute names to find country). Also manually we merged the universities and departments which had different names (e.g., misspellings or using abbreviations) in our original extracted.

After the cleansing of the publication data, the resulting database contained 2,226 papers reflecting the contributions of 5,201 authors (117 of authors had no affiliation data that we ignored them in our analysis) from 1,324 institutes (i.e., universities and private companies) of 76 countries. We found 447 multi-affiliation authors who for 411 of them just two affiliations have been recorded. While it is possible to have different authors with similar academic names, but since we are focusing on a specific discipline, it is more reasonable to regard the ones that have similar academic name as an author with multi-affiliations. To validate that we checked manually for some of them and find them as an author who was graduated in one institute and then has been worked as researcher or a professor in other institutes.

Figure 1 depicts the development and evolution of publications in "steel structures" research by showing frequency of publications per year between 1970 and 2009. The timely evolution of publications is marked for each year. The first paper (based on our dataset) was published in 1970 and it was the only publication at that year, second one was published in 1972 and then a few publications (minimum one and maximum five) per year until 1980 with a jump to 15 publications in 1981. The number of publications had a huge growth in 1983 46 publications. The data shows a moderate increase with minor fluctuations between 1984 and 1997 but a strong take off in the 1998. For instance, the number of 76 publications in 1997 increased to more than 300 in 1998 and again a sharp decline to about 120 in 1999. The huge peak of publications in 1998 is because of publishing abstracts of a special conference (held in that year) in special series in one of the journals. While the data after this big jump shows fluctuations but seem to confirm the continuation of this development by a steady increase of number of publications over time overall.

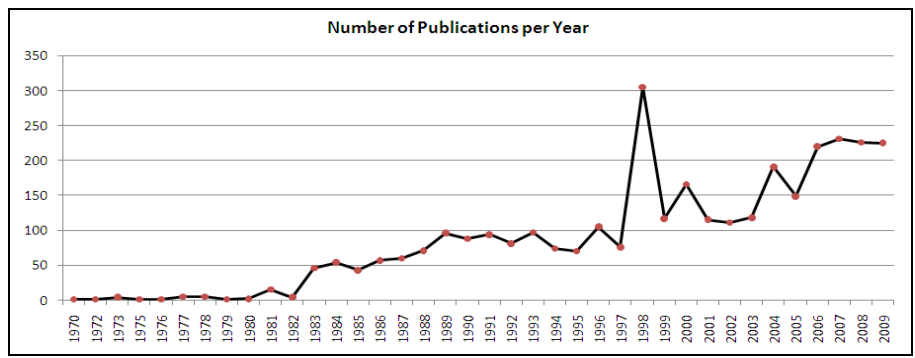

Figure 1. Frequency of Published Papers per year

Although the publications dataset that we have extracted are not representative of the world production in the "steel structures" field (as there is the possibility of significant biases: evolving list of relevant journals, publications in other journals, same field using other words in a 40 years period) but the database does not pretend to represent the field, and is just to illustrate the use of the indicators to analyse this field. Therefore, the biases are not a problem for our analysis.

## 3. Analysis of Collaboration Networks

Figure 2 (a) shows the frequency of publications against the number of authors. It follows that most publications in "steel structures" are written by two authors (41%) followed by 3-author and single-author publications (25% and 23% respectively). The low percentage of single-author publications (23%) demonstrates a high collaborative attitude in this field. The distributions shown in Figures 2 (b-c) indicate that the majority of publications are result of single- or multi-authors nips from single institutes (70%) and single countries (87%), although numerous inter-institutional and international collaborations are also observed.

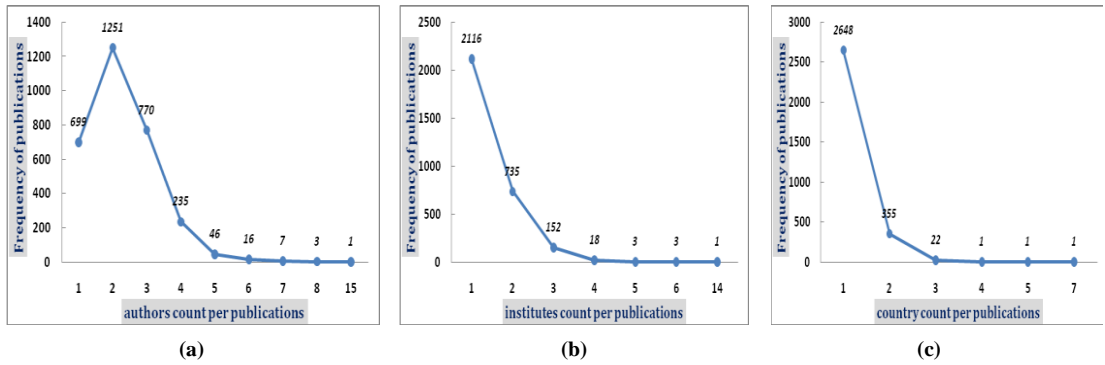

Figure 2: distributions of number of publications and different level of affiliations

In the following sections, the most collaborative countries, institutes and authors in "steel structures" are identified by the origin of the authors. We consider the number of publications which have authors from different institutes or countries (at least one author from a different institute or country) as an indication of inter-institutional or international collaboration. For instance, if a publication has four authors, of whom one is from an institute (or country) different from the others, it is considered as an inter-institute (or international) collaboration.

### 3.1. Macro-Level Analysis

The distribution of international collaborations over the last 40 years is now analyzed in order to identify the countries collaboration activity. Figure 3 shows the evolution with time of international collaborations (macro-level) on "steel structures" research including the number of distinct collaborators (collaborative countries), frequency of unique collaboration links (number of links) and sum of collaborations (sum of weight of the links) per year. The first international collaboration appears in 1972 in a paper that had at least one author from United States and China. The overall trend shows increasing amount of collaborations over the time while there are few fluctuations in some period of times.

Totally 1,076 international collaborations for 188 links (unique pair of countries) among 66 countries (nodes) have been extracted from our dataset. The most collaborative year was 2006 with 192 collaborations for 51 links among 27 unique collaborative countries. The trend shows the number of countries per year remains almost steady after 2005 between 25 and 27 (out of 66) countries that have at least one international collaborations. This indicates that just few countries are active and having publications with international collaborations almost every year and the remaining had just few publications with international collaborations just once or twice over 40 years.

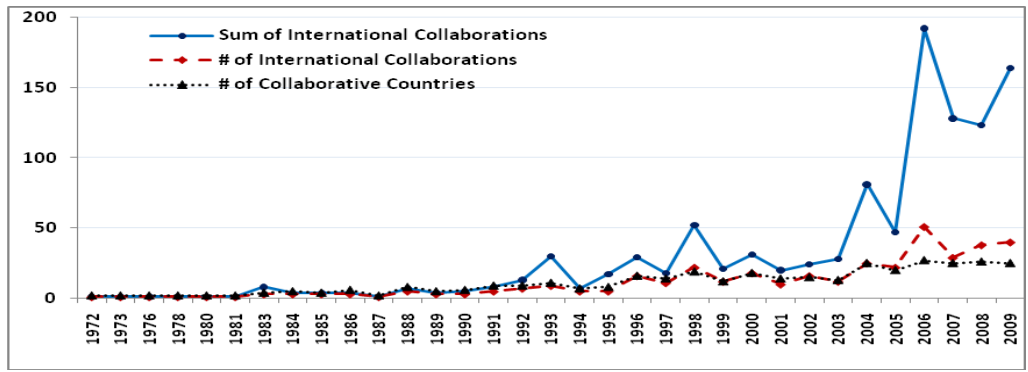

Figure 3. Frequency of international collaborations per year

Figure 3 depicts almost similar trend for number of collaborations and sum of collaborations per year except for 2008 that while number of international collaborations is higher than 2007 but sum of collaborations for 2008 is lower than 2007. This shows on average weaker links (less collaboration) between each pair of countries in 2008 compare to 2007. The big gaps between number of collaborations and sum of collaborations after 2005 could be explained due to having more collaboration between each pair of collaborating countries (each link) compare to 2005 and before.

### 3.1.1. Identifying the Top Collaborating countries and Strong Links

Table 1 shows the top collaborating countries that have six or more international collaborations. They are listed in descending order of sum of international collaborations (# Col) followed by the number of links to the collaborating countries (# Cnt).

Table 1. Top collaborative countries: country name, sum of collaborations (# Col) and number of collaborators (# Cnt).

|    | Country        | # Col | # Cnt |    | Country              | # Col | # Cnt |
|----|----------------|-------|-------|----|----------------------|-------|-------|
| 1  | United States  | 413   | 36    | 21 | Egypt                | 19    | 4     |
| 2  | China          | 297   | 13    | 22 | Iran                 | 15    | 5     |
| 3  | United Kingdom | 167   | 32    | 23 | Turkey               | 13    | 6     |
| 4  | Australia      | 152   | 19    | 24 | India                | 13    | 6     |
| 5  | Canada         | 116   | 16    | 25 | New Zealand          | 13    | 2     |
| 6  | Japan          | 107   | 12    | 26 | Bangladesh           | 12    | 4     |
| 7  | France         | 84    | 18    | 27 | Slovakia             | 11    | 4     |
| 8  | Germany        | 80    | 17    | 28 | South Africa         | 11    | 4     |
| 9  | South Korea    | 75    | 8     | 29 | Malaysia             | 10    | 4     |
| 10 | Singapore      | 47    | 8     | 30 | Sweden               | 9     | 5     |
| 11 | Czech Republic | 47    | 9     | 31 | Viet Nam             | 9     | 3     |
| 12 | Belgium        | 47    | 11    | 32 | Ireland              | 8     | 2     |
| 13 | Portugal       | 46    | 10    | 33 | Serbia               | 8     | 3     |
| 14 | Italy          | 45    | 8     | 34 | Cyprus               | 8     | 2     |
| 15 | Netherlands    | 43    | 11    | 35 | Norway               | 8     | 2     |
| 16 | Finland        | 30    | 8     | 36 | United Arab Emirates | 8     | 3     |
| 17 | Luxembourg     | 27    | 6     | 37 | Brazil               | 7     | 3     |
| 18 | Spain          | 23    | 9     | 38 | Poland               | 7     | 4     |

| | | | | | | |
|---|---|---|---|---|---|---|
| 19 | Greece | 21 | 7 | 39 | Romania | 6 | 3 |
| 20 | Switzerland | 19 | 11 | 40 | Jordan | 6 | 3 |

It follows from Table 1 that United States is the most collaborative country in this field having a total of 413 international collaborations. In second position, China (including Hong Kong and Taiwan) has 297 collaborations followed by United Kingdom, Australia and Canada with 167, 152, 116 collaborations respectively. In terms of collaborating countries, the ranking is different and United States and United Kingdom have the highest number of collaborators followed by Australia, France and Germany (36, 32, 19, 18 and 17 collaborators respectively).

In order to identify the strongest collaborations (links), we evaluated the frequency of collaboration between each pair of countries. Table 2 shows the top 20 strongest international collaborations. China and United States have the most frequent collaboration (link) with 125 collaborations over the last 40 years followed by the link between China and Australia with 69 collaborations. Interestingly, the top 20 links (out of 188), among 18 countries (out of 76), contains more than half of the international collaborations (570 out of 1,076). In another words, about 11% of the international collaboration links include more than 50% of the international collaborations.

Table 2. Top strongest international collaborations: country, co-country and sum of collaborations (# Col.).

|  | Country | Co-Country | # Col. |  | Country | Co-Country | # Col. |
|---|---|---|---|---|---|---|---|
| 1 | China | United States | 125 | 11 | China | Singapore | 13 |
| 2 | China | Australia | 69 | 12 | France | Germany | 12 |
| 3 | Japan | United States | 57 | 13 | Germany | United Kingdom | 12 |
| 4 | United States | South Korea | 52 | 14 | Netherlands | Belgium | 12 |
| 5 | Canada | United States | 46 | 15 | New Zealand | United States | 12 |
| 6 | Japan | China | 34 | 16 | United Kingdom | Portugal | 12 |
| 7 | China | United Kingdom | 25 | 17 | Belgium | Luxembourg | 11 |
| 8 | Canada | Australia | 16 | 18 | France | United Kingdom | 11 |
| 9 | Czech Republic | United Kingdom | 15 | 19 | Italy | United States | 11 |
| 10 | United States | Germany | 14 | 20 | Singapore | United States | 11 |

### *3.1.2. Analyzing the Network of Collaboration Countries*

From total 76 countries which had at least one publication, 10 countries (*Albania; Bulgaria; Croatia; Estonia; Iraq; Latvia; Lithuania; Republic of Macedonia; Ukraine; Venezuela*) did not have any international collaboration while even some of them like Lithuania and Ukraine have 12 and 7 publications respectively. The network of 1,076 international collaborations with 188 links (pair of countries) has been analyzed using UCINET (Borgatti, Everett et al. 2002). Table 3 shows results of the analysis.

Table 3. International collaboration network measures

| Measures | Values | Measures | Values (%) |
|---|---|---|---|
| **Density** | 50.2% | **Average distance** | 2.38 |
| **Connectedness** | 100% | **Centralization Measures** | |
| **Clustering Coefficients** | 8.16% | Degree | 4.75 |
| **# of Components** | 1 | Closeness | 61.82 |
| **Giant Component Size** | 66 | Betweenness | 34.74 |

The network can be considered as a dense network as almost half of the possible links among all nodes (countries) exist. The network is fully connected and all nodes belong to a component (a group of nodes that are all connected to one another by paths of intermediate acquaintances) whereby the whole network is the "giant component". The clustering coefficients value for the network shows that the probability that any two unconnected collaborators of a country, collaborates (wrote a paper) together is just 8% which is very low. The international network is decentralized considering the degree centrality of the nodes but the network is centralized around a few countries close to all others on average (having high closeness centrality). Average distance among any country to reach any other one in the network is 2.38 meaning that just by two steps (on average) each country reaches another one.

Figure 4 shows the international collaboration network of "steel structures" research. Different widths of links reflect different frequency of collaboration links (the thicker the link, the more collaboration between the two connected nodes). As shown in the graph, United States, United Kingdom, Australia, France, Germany and Canada are the most connected.

Different nodes' color and shape distinguished the regions of countries (blue rectangles for America (including North and Latin America) on the up-right side of the figure; red diamond for Oceania on the right side of the figure; blue hexagonal for Africa on the bottom-right side of the figure; red circle for Asia on the up-left side of the figure; and purple triangle for Europe on the bottom-left side of the figure). It appears that, there is much inter-region collaboration between America and European and Asia. The most internal collaborations exist among European countries.

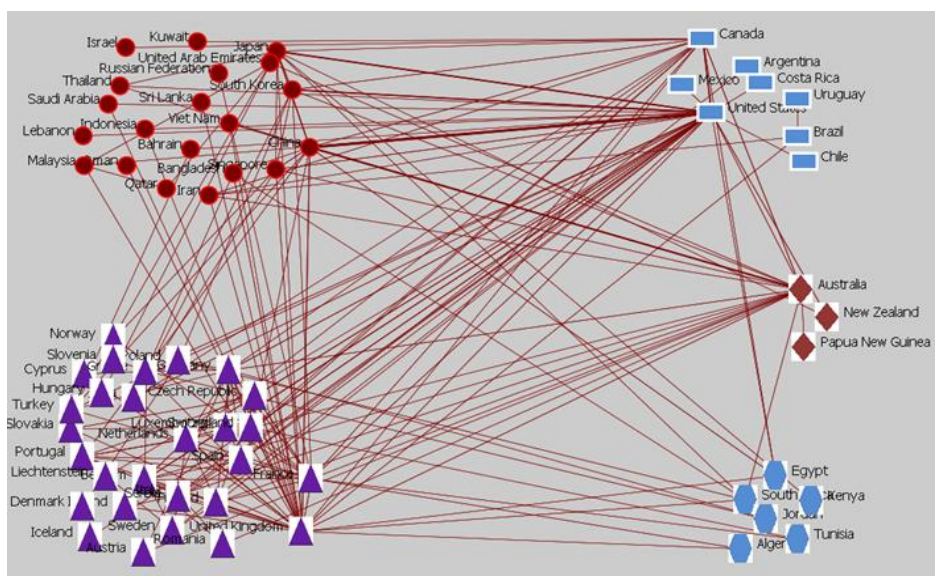

Figure 4. International collaboration network (1970-2009)

We extracted the cliques of the international collaboration network and found 62 cliques with at least 3 nodes. A clique is a subgroup in which all its nodes are directly connected to each other (while a cluster is a group of the same or similar elements gathered or occurring closely together). The largest size of clique is 7 and out of total 62 cliques, just 2 cliques have 7 nodes or more. It means that in each clique of 7 nodes, each of one is connected to 6 other nodes). The lists of countries in the biggest cliques of international network are:

> 1: *United States; Australia; Canada; United Kingdom; Germany; France; Switzerland*
> 2: *United States; United Kingdom; Czech Republic; Germany; Netherland; France; Finland*

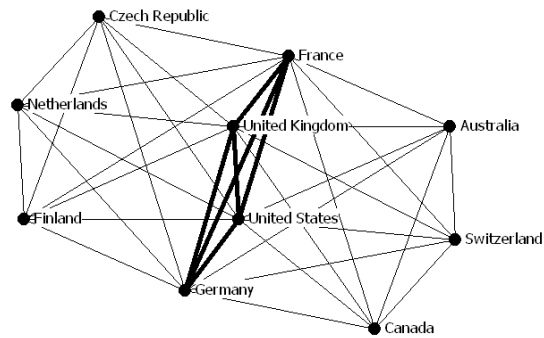

Figure 5. International collaboration among biggest cliques in network

### 3.2. Meso- Level Analysis

Here we focus on collaborations between institutes (e.g., university, research institutes and corporations). Figure 6 shows the evolution over time of inter-institutional collaborations (meso-level) for the "steel structures" field having distinct collaborators (collaborative institutes), number of unique collaborations (links) and sum of collaborations per year. The trends for the three measures are similar except that the number of collaborators which have an almost fixed value after 2006 (similar to the marco-level) and that the value for the number of collaborations are equal or slightly lower than sum of collaborations. This indicates general weak links among institutes which means basically one (or few) paper per link. The overall trend shows slightly increasing amount of collaborations over the time having some fluctuations and large increases in 2004 and 2006. The most collaborative year is 2009 with 379 collaborations for 191 links among 170 unique collaborators (institutes) but in terms of number of links, 2006 is ranked first with 224 links which indicates stronger links among collaborative institutes (more than one paper per link) in 2006.

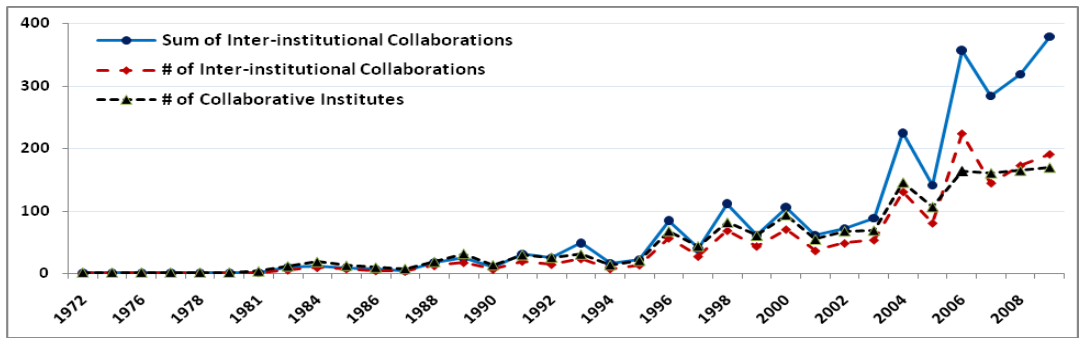

Figure 6. Frequency of inter-institutional collaborations per year

### 3.2.1. Identifying the Top Collaborative Institutes and Strong Links

Evaluating the 2,583 collaborations for 1,514 links among 907 intuitions, we ranked the institutes in terms of total collaborations they have. Table 4 shows the top collaborative institutes in the "steel structures" field and their sum of collaborations and number of collaborators (number of unique institutes which have at least one co-authored paper with the institute. "*Tsinghua University*" from China is ranked first followed by "*Imperial College London*" from United Kingdom and "*University of Sydney*" from Australia with 115, 82 and 65 inter-institute collaborations respectively over the last 40 years.

Table 4. Top collaborative institutes: institute name, country, country, sum of collaborations (# Col) and number of collaborators (# Cnt).

|    | Institute | Country | # Col | # Cnt |
|----|-----------|---------|-------|-------|
| 1  | Tsinghua University | China | 115 | 33 |
| 2  | Imperial College London | United Kingdom | 62 | 24 |
| 3  | University of Sydney | Australia | 56 | 25 |
| 4  | Fuzhou University | China | 54 | 10 |
| 5  | University at Buffalo | United States | 36 | 17 |
| 6  | Monash University | Australia | 35 | 19 |
| 7  | Purdue University | United States | 28 | 17 |
| 8  | Ecole Polytechnique de Montreal | Canada | 27 | 18 |
| 9  | University of Minnesota | United States | 27 | 18 |
| 10 | Hong Kong Polytechnic University | China | 27 | 18 |
| 11 | University of Coimbra | Portugal | 24 | 15 |
| 12 | University of Manchester | United Kingdom | 24 | 22 |
| 13 | University of California at Davis | United States | 23 | 17 |
| 14 | University of Illinois at Urbana-Champaign | United States | 22 | 18 |
| 15 | Georgia Institute of Technology | United States | 21 | 11 |
| 16 | University of New South Wales | Australia | 19 | 12 |
| 17 | Louisiana State University | United States | 18 | 10 |
| 18 | University of British Columbia | Canada | 18 | 14 |
| 19 | Lehigh University | United States | 18 | 12 |
| 20 | University of Alberta | Canada | 18 | 14 |

Interestingly while United States is the most collaborative country but the first institute from United States in the list is ranked 5[th] and there are just three institutes from United States in top 10 collaborative institutes. This shows that overall huge numbers of institute are active in this field in United States but none of them (individually) is not in top 4 collaborative institutes. On the other hand, "*University of Coimbra*" from Portugal is ranked 11[th] but Portugal is not among in the top 20 collaborative countries.

Table 5 shows the top 10 strongest inter-institutional collaborations links among institutes which have nine or more collaborations. Two Chinese universities have the strongest inter-institutional collaborations followed by two universities from Brazil and Italy. Among the 10 collaborations, two Chinese institutes have international collaborations with an institute in US and Egypt. Thus, data shows 80% of the top 10 strong collaboration links are intra-national collaborations (i.e., two institutes are located in the same country). The data also shows that most of the institutes in this table are not among the top collaborative institutes in Table 4 which means that the institutes listed in Table 5 have strong links with only a few number of institutes (collaborators).

Table 5. Top strongest inter-institutional collaborations: institute name, country and sum of collaborations (# Col).

|    | Institutes                   | Country       | Co-Institutions                          | Co-Country    | # Col |
|----|------------------------------|---------------|------------------------------------------|---------------|-------|
| 1  | Fuzhou University            | China         | Tsinghua University                      | China         | 35    |
| 2  | State Uni of Rio De Janeiro  | Brazil        | Pontifical Catholic Uni of Rio de Janeiro | Brazil        | 31    |
| 3  | University of Chieti         | Italy         | University of Naples Federico II         | Italy         | 21    |
| 4  | Ecole des Mines de Paris     | France        | EDF Center des Renardi ãres              | France        | 11    |
| 5  | National Taiwan University   | China         | Nat'l Cent for Res on Earthquake Eng     | China         | 10    |
| 6  | University of Minnesota      | United States | Georgia Institute of Technology          | United States | 10    |
| 7  | Zhejiang University          | China         | CSCEC Nat'l Swim Cent Design Cons        | China         | 10    |
| 8  | University of Hong Kong      | China         | Tanta University                         | Egypt         | 9     |
| 9  | University of New South Wales | Australia    | Monash University                        | Australia     | 9     |
| 10 | Kyoto University             | United States | Tsinghua University                      | China         | 7     |

### *3.2.2. Analyzing the Network of Collaborating Institutes*

Of the 1,342 institutes which had at least one publication, only 907 (about 68%) institutes have at least an inter-institutional collaboration. By analyzing this reduced network with 907 nodes and 1,514 links, the network structure measures shown in Table 6 have been obtained. It follows from Table 6 that, the network is very sparse with density of less than 1% consists of 98 components (with different sizes from 2 to 9 and 649) are found for institutional networks and the largest component contains 649 connected nodes (institutes) and the second largest component are 9 connected institutes, which is far smaller than the largest one and it also verifies that the inter-institutional scientific collaboration networks are not on the borderline of connectedness (Newman 2001). Only about 51% of the institutes are reachable by each other (there is a link with any length among each two nodes).

Table 6. Inter-institutional collaboration network measures

| Measures | Values | Measures | Values (%) |
|---|---|---|---|
| **Density** | 0.63% | **Average distance** | 4.97 |
| **Connectedness** | 51.2% | **Centralization Measures** | |
| **Clustering Coefficients** | 1.4% | Degree | 0.35 |
| **# of Components** | 98 | Closeness | 29.05 |
| **Giant Component Size:** | 649 | Betweenness | 10.02 |

The clustering coefficients for the network demonstrate a very low probability that any two collaborators of an institute are collaborating. The inter-institutional network may be described as a ring network as its degree centralization value is very low but the network is somehow centralized around a few institutes having high closeness centrality. By five steps (on average) an institutes can reach any other institutes in their collaboration network.

Figure 7 shows a 3D view of the institutional collaboration network. It shows a large connected subgroup (the giant component) in the bottom part of the figure and several other unconnected subgroups above that. Due to the large size of the network, we cannot distinguish particular nodes (institutes) but the Figure provides an overview of the network.

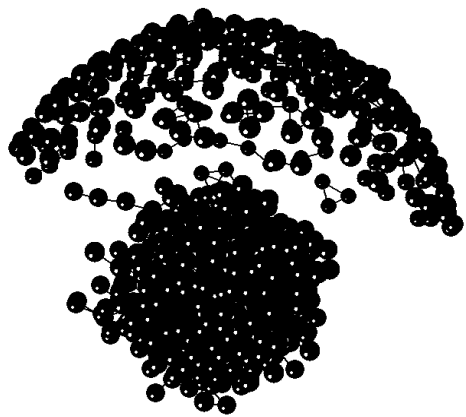

Figure 7. Inter-institutional collaboration network

248 cliques have been found with at least three directly connect nodes which the largest clique contains 14 fully connected institutes (which each of them is connected to other 13 institutes in the clique) followed by three other cliques with size of 6. The cliques and their member are listed below and Figure 8 illustrates that these four largest cliques in inter-institutional network are fragmented except clique number 2 (mainly Canadian institutes) and number 3 (mainly Italian institutes). Only one institute is part of two cliques, namely the "*École Polytechnique de Montréal*", which connects members of each clique to the other one.

1: *University of Coimbra; EPFL; Czech Technical University; Slovak Technical University; British Constructional Steelwork Association; Building Research Establishment Ltd* 2: *École Polytechnique de Montréal; University of British Columbia; Groupe RSW; TVP Engineering; University of Sherbrooke; Campus de Clermont-Ferrand-Les Cézeaux*
3: *École Polytechnique de Montréal; University of Pisa; University of Trento; Corporate Research Policies of Riva Group S.p.A.; IPSC-ELSA; University of Molise*
4: *University of Manchester; EDF Center des Renardières; NRI; VTT; AREVA NP SAS; CEA; AREVA NP GmbH; E.ON Energie; IWM; University of Stuttgart' EC JRC-IE; British Nuclear Group; Serco Assurance; ORNL*

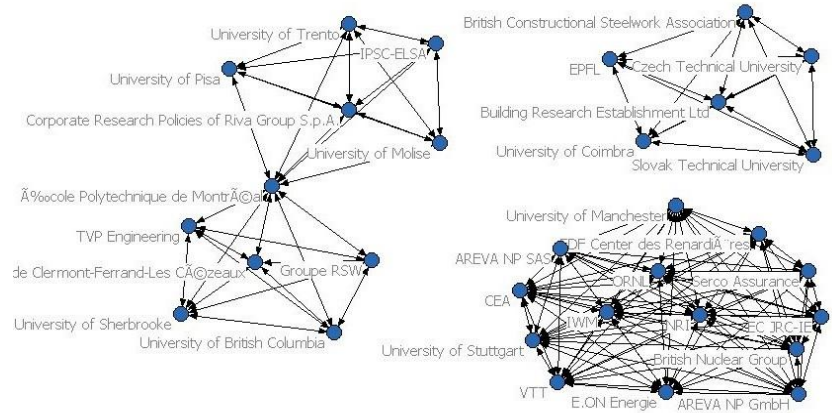

Figure 8. Inter-institutional collaboration among biggest cliques in network

### 3.3. Micro-Level Analysis

Figure 9 shows the evolution with time of co-authorship network of "steel structures" researchers, depicting the number of authors (nodes) having at least one collaboration, the number of collaborations (links) and sum of collaborations (weight of links) per year over the last 40 years. The trend for number of collaborations and sum of collaborations is almost the same just by having equal or slightly lower value for number of collaborations compare to sum of collaborations. This indicates general weak links among authors rather (having one or very few co-authored papers) than few but strong links.

As shown, numbers of authors with at least one collaboration (per year) are lower than number and sum of collaborations and the gap among them is more after 2005, due to having more redundant collaborations (more than one paper) between co-authors. On the other hand, numbers of authors are higher than number and sum of collaborations between 1970 and 1992 which indicates mostly authors have one (or very few) publication with one co-author per year.

The overall trend shows slightly increasing amount of collaborations over the time. The gap between numbers of authors and number and sum of collaborations increases after 2006 due to having stronger links (more collaboration). The number and sum of collaborations and number of authors have been raised dramatically in 1998 following with a quick big decrease. The huge peak of publications in 1998 is because of publishing abstracts of a special conference held in that year in a special series in one of the journals. The same fluctuation has been happened in 2006. The most collaborative year is 2009 with 695 collaborations (links) and totally 749 collaborations between 512 unique authors.

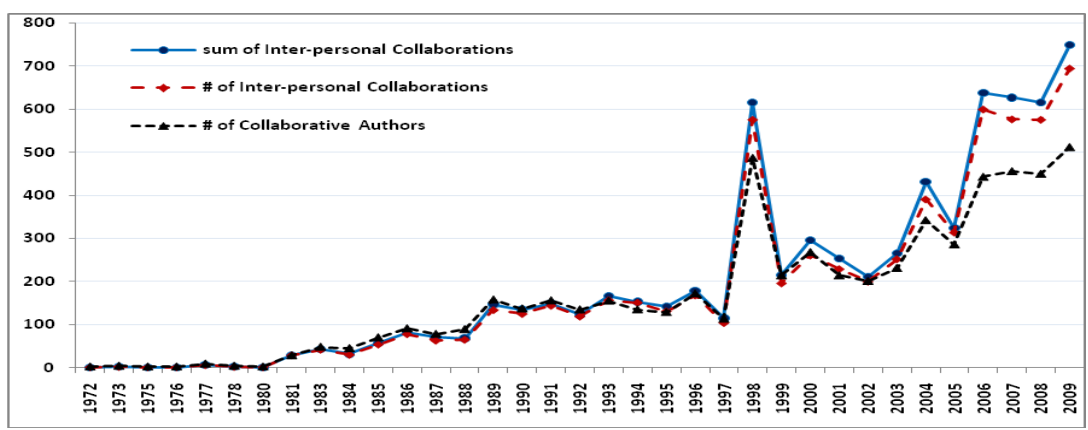

Figure 9. Frequency of co-authorship collaborations per year

### *3.3.1.    Identifying the Top Collaborative Authors and Strong Links*

Table 7 shows the top collaborative authors in this field of study, each with more than 28 collaborations. They are in order of the total number of collaborations but numbers of co-authors (number of unique authors which have at least one co-authored publication) is also shown for each author. Although "*L.-H. Han*", from Tsinghua University is ranked first followed by "*B. Young*" from University of Hong Kong, but "*Nakashiman*" from Kyoto University has more collaborators.

Table 7. Top collaborative authors: author name, institute name, country, sum of collaborations (# Col) and number of co-authors (# Cnt).

|    | Author          | Institute                          | Country        | # Col | # Cnt |
|----|-----------------|------------------------------------|----------------|-------|-------|
| 1  | L.-H. Han       | Tsinghua University                | China          | 66    | 26    |
| 2  | B. Young        | University of Hong Kong            | China          | 61    | 19    |
| 3  | M. Nakashiman   | Kyoto University                   | Japan          | 59    | 42    |
| 4  | G.J. Hancock    | University of Sydney               | Australia      | 57    | 27    |
| 5  | W.-F. Chen      | National University of Singapore   | Singapore      | 55    | 33    |
| 6  | I.W. Burgess    | University of Sheffield            | United Kingdom | 53    | 17    |
| 7  | D.A. Nethercot  | Imperial College London            | United Kingdom | 52    | 28    |
| 8  | X.-L. Zhao      | Monash University                  | Australia      | 49    | 25    |
| 9  | R.J. Plank      | University of Sheffield            | United Kingdom | 47    | 15    |
| 10 | N.E. Shanmugam  | National University of Singapore   | Singapore      | 42    | 22    |
| 11 | K.J.R. Rasmussen| University of Sydney               | Australia      | 41    | 20    |
| 12 | R. Tremblay     | Ecole Polytechnique de Montreal    | Canada         | 36    | 24    |
| 13 | T. Usami        | Nagoya University                  | Japan          | 35    | 14    |
| 14 | J.M. Ricles     | Lehigh University                  | United States  | 33    | 13    |
| 15 | Z. Tao          | Fuzhou University                  | China          | 31    | 18    |
| 16 | S.L. Chan       | Hong Kong Polytechnic University   | China          | 30    | 19    |
| 17 | J.M. Rotter     | University of Edinburgh            | United Kingdom | 29    | 17    |
| 18 | A.S.A.L. de     | State University of Rio De Janeiro | Brazil         | 28    | 14    |
| 19 | K.H. Tan        | Nanyang Technological University   | Singapore      | 28    | 14    |

Table 8 shows the top strongest co-authors collaborations with seven or more collaborations (joint-publications). The strongest co-authorship collaborations are among two authors from the "*University of Sheffield*" following by two authors from the "*University of Sydney*" and two authors from "*Nagoya University*". As the data shows, only five of the top co-authorship collaborations (about 34%) are inter-institutional collaborations and only three (20%) are international collaborations. High intra-institutional collaborations indicate that the co-authors may have advisor-student relationship. The data also shows that most of the authors are not among the top collaborative authors.

Table 8. Top strongest inter-institutional collaborations: institute name, country and sum of collaborations (# Col).

|    | Authors      | Institute                         | Co-author       | Co-institutions                   | # Col |
|----|--------------|-----------------------------------|-----------------|-----------------------------------|-------|
| 1  | I.W. Burgess | University of Sheffield           | R.J. Plank      | University of Sheffield           | 20    |
| 2  | G.J. Hancock | University of Sydney              | K.J.R. Rasmussen| University of Sydney              | 11    |
| 3  | H. Ge        | Nagoya University                 | T. Usami        | Nagoya University                 | 10    |
| 4  | L.-H. Han    | Tsinghua University               | Z. Tao          | Fuzhou University                 | 10    |
| 5  | B. Young     | University of Hong Kong           | E. Ellobody     | Tanta University                  | 9     |
| 6  | B. Young     | University of Hong Kong           | F. Zhou         | University of Hong Kong           | 9     |
| 7  | D. Camotim   | Institute Superior Tecnico        | N. Silvestre    | Institute Superior Tecnico        | 8     |
| 8  | J.M. Ricles  | Lehigh University                 | R. Sause        | Lehigh University                 | 8     |
| 9  | Y.-L. Pi     | University of Sydney              | N.S. Trahair    | University of Sydney              | 8     |
| 10 | B. Ahmed     | Bangladesh Univ of Eng. & Tech.   | D.A. Nethercot  | Imperial College London           | 7     |
| 11 | B. Young     | University of Hong Kong           | J. Chen         | University of Hong Kong           | 7     |
| 12 | K. Suita     | Kyoto University                  | M. Nakashiman   | Kyoto University                  | 7     |
| 13 | L.-H. Han    | Tsinghua University               | X.-L. Zhao      | Monash University                 | 7     |
| 14 | L.-H. Han    | Tsinghua University               | Y.-F. Yang      | Fuzhou University                 | 7     |
| 15 | S.L. Lee     | National University of Singapore  | N.E. Shanmugam  | National University of Singapore  | 7     |

### *3.3.2. Analyzing the Network of Collaborating Authors*

6,111 unique links (pair of co-authors) with totally 7,386 collaborations are extracted from the database. We omitted showing the network diagram due to the large number of nodes and links. Some network structure measures shown in Table 9. The co-authorship network is very sparse as about 15% of authors are reachable. As it is shown, 753 components with different sizes (2-16, 18-21, 23, 28 and 1561 nodes) are found for institute networks and the largest component size (the giant component) is 1561 and the second largest group of connected authors is 28 which is again far smaller than the largest one. The network is decentralized considering degree and betweenness centralities of the nodes but the network is more centralized around few countries close to all others on average. The average distance (among reachable pairs) is 7.90 which is the average number of links in the network, between all pair of scholars for whom a connection exist. So, the typical distance between each pair of scholars is 8 for "steel structures" researchers.

Table 9. Co-authorship collaboration network measures

| Measures | Values | Measures | Values (%) |
|---|---|---|---|
| Density | 0.09% | Average distance | 7.90 |
| Connectedness | 14.7% | Centralization Measures | |
| Clustering Coefficients | 1.17% | Degree | 0.08 |
| # of Components | 753 | Closeness | 14.14 |
| Giant Component Size: | 1561 | Betweenness | 3.1 |

We extract 1,116 cliques with at least 3 nodes. The largest cliques includes one clique of 15 fully connected authors, one clique of 9 authors and two cliques including 8 fully interconnected authors (to all other 7 authors) as shows below. As shown in Figure 10, none of the authors from above cliques are connected to other cliques that lead cliques to be fragmented from each other.

*1: A. Ghobarah D. Mitchell R. Tinawi M. Saatcioglu A.G. Gillies D. Lau D.L. Anderson N.J. Gardner*
*2: D.B. Moore T. Lennon A. Santiago da Silva L. Simoes F. Wald L. Borges M. Benes M. Chladna*
*3: A. Elgamal B. Moaveni D.H. Whang F. Tasbihgoo J.P. Conte M. Wahbeh S.F. Masri X. He*
*4: A.H. Sherry B.R. Bass D. Connors D. Lauerova D. Siegele D.P.G. Lidbury E. Keim G. Nagel H. Keinanen K. Nilsson K. Wallin P. Gilles S. Marie U. Eisele Y. Wadier*

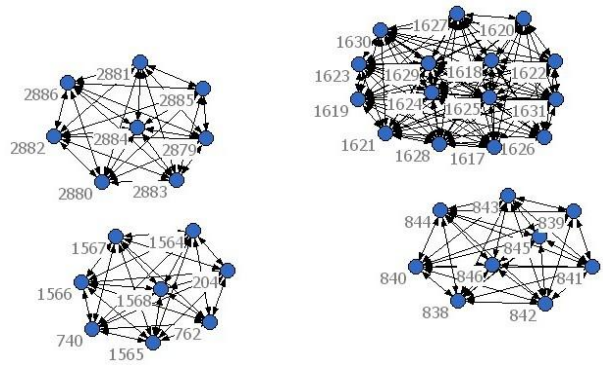

Figure 10. Co-authorship collaboration among biggest cliques in network

## 4. Dynamics of Nations: Cross-time Analysis of Growth Trends of Countries

Looking at Figure 1, during the evolution of "steel structures" research's publication there is two important points: first there is sharp increase in 1983 and 1998 (after a steady increase during 1970-1982 and 1983-1997. On the other hand, the Scopus-listed publications pre-1983 is very small while the pre-1983 frequency of publication was higher than shown in Figure 1 due to the fact that the electronic publication of research on "steel structures" started around 1982 and hence. Therefore, we will have a macro-level cross-time analysis for the two periods: 1983-1997 and 1998-2009.

### 4.1. Growth Trend of Top Collaborative Countries

Table 10 shows the top 20 collaborating countries (in order of the sum of collaboration) for each three periods of times. There were only 33 countries during 1983-1997 that had at least one collaboration through 50 distinct pairs of collaborations (links) over the 15 years but 163 pairs of collaborations have extracted among 55 countries for the following 11 years (during 1998-2009). In addition, sum of collaborations had a sharp increase from 159 during 1983-1997 to 911 during 1998-2009.

Studying the evolution of the scientific collaboration activity amongst countries, United States has the majority of collaborations followed by China and United. Interestingly France, South Korea, Singapore and Portugal among are the top 10 collaborating countries in the second period (1998-2009), but were not in the top 10 ranked countries in the first period.

Table 10. Top 20 collaborative countries: country name, sum of collaborations (# Col) and number of collaborating countries (# Cnt).

|   | 1983 - 1997 | | | 1998 – 2009 | | |
|---|---|---|---|---|---|---|
|   | Country | # Col | # Cnt | Country | # Col | # Cnt |
| 1 | United States | 60 | 11 | United States | 350 | 34 |
| 2 | China | 31 | 5 | China | 265 | 12 |
| 3 | United Kingdom | 30 | 12 | United Kingdom | 136 | 28 |
| 4 | Japan | 25 | 6 | Australia | 133 | 19 |
| 5 | Luxembourg | 24 | 4 | Canada | 99 | 13 |
| 6 | Australia | 18 | 6 | France | 81 | 17 |
| 7 | Canada | 16 | 5 | Japan | 81 | 7 |
| 8 | Belgium | 15 | 4 | South Korea | 70 | 7 |
| 9 | Germany | 12 | 5 | Germany | 68 | 13 |
| 10 | Greece | 12 | 3 | Portugal | 45 | 10 |
| 11 | Singapore | 11 | 2 | Czech Republic | 44 | 9 |
| 12 | Netherlands | 8 | 1 | Italy | 38 | 7 |
| 13 | Italy | 7 | 4 | Singapore | 36 | 8 |
| 14 | Jordan | 6 | 3 | Netherlands | 34 | 9 |
| 15 | Turkey | 5 | 2 | Belgium | 32 | 10 |
| 16 | South Korea | 5 | 3 | Finland | 30 | 8 |
| 17 | Iran | 4 | 3 | Spain | 21 | 8 |
| 18 | India | 3 | 2 | Switzerland | 18 | 11 |
| 19 | France | 3 | 2 | Egypt | 17 | 4 |
| 20 | Czech Republic | 2 | 1 | New Zealand | 13 | 2 |
| **Sum of Links:** | | | **159** | | | **911** |
| **Number of Links:** | | | **50** | | | **163** |
| **Number of Countries:** | | | **33** | | | **59** |

Figures 11 and 12 show the international collaboration network of "steel structures" field for each period. Different nodes' shape distinguish the regions of countries (rectangles for America; diamond for Oceania; hexagonal for Africa; circle for Asia; and triangle for Europe).

It follows from Figure 11 that during the 1983-1998 period, Japan and United States had the largest number of international collaboration link followed by strong collaboration between United States and China. During this period, no county from Latin America and only one country from Africa had international collaboration. Although there are good number of collaborating countries from Asia but with very weak intra-region links. European countries not only had collaborations with other regions but also had a strong collaboration among them especially between Luxembourg and Belgium and Netherlands.

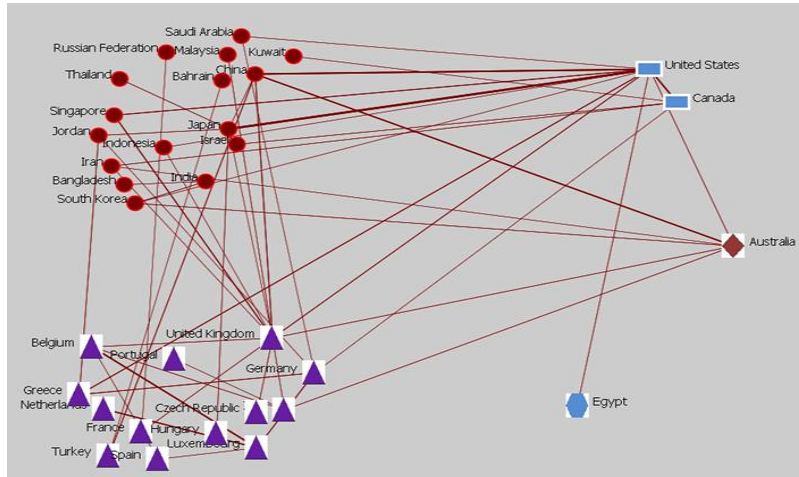

Figure 11. International collaboration network during 1983-1997

More countries and international collaborations have been depicted during 1998-2009 in Figure 12 on the field of "steel structures". China and United States had the strongest link (112 collaborations) followed by China and Australia and South Korea and United States. New collaborations have been recorded for the first time not only for Latin American and African (except Egypt) countries but also some other countries in Asia, Europe and Oceania in this period as follows from Figure 12. Interestingly, no intra-region collaborations exist among Latin American and African countries. Similar to previous periods, European countries had the most intra-region collaborations.

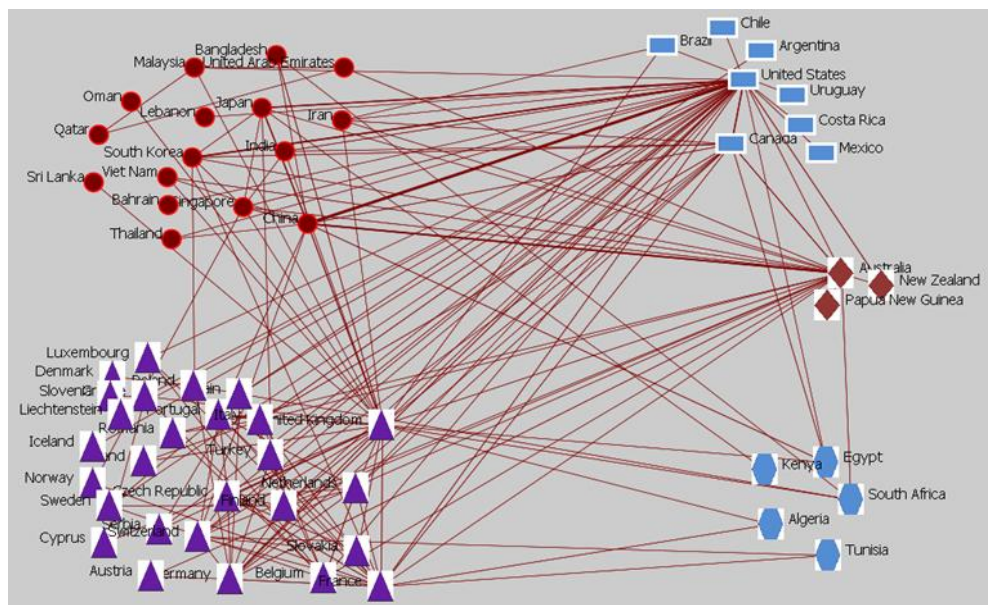

Figure 12. International collaboration network during 1998-2009

Table 11 compares the network measures of the international collaborations of the "steel structures" over time for two specific periods. The results show only one components exist (all countries are reachable) but the "giant component" size has increased as the network becomes denser over time. Thus, the international collaboration network is not fragmented in many components and all nations belong to a single giant component in which all nodes are connected to one another by paths of intermediate acquaintances. The increasing in clustering coefficients over time means that it is significantly common for nations to broker new collaborations. Interestingly, while the international network during 1998-2009 has more nodes and links than the 1983-1997 period, the average path between countries was lower.

Considering network centralization measures to explore the structural change of the international network, as the size of the network increases, its degree centralization measures decreases while the closeness and betweenness centralization measures are increasing. This means that the network structure is becoming centralized around a few countries close to all other countries (having high closeness) and brokering more between others (high betweenness centrality).

Table 11. Cross-time international collaboration network measures

| Measures | 1970-09 | 1983-97 | 19980-09 |
|---|---|---|---|
| Density | 50.2 | 30.11 | 53.24 |
| Connectedness | 100 | 100 | 100 |
| Clustering Coefficients | 8.16 | 1.41 | 8.18 |
| # of Components | 1 | 1 | 1 |
| Giant Component Size: | 66 | 33 | 59 |
| Average Distance | 2.39 | 2.90 | 2.36 |
| **Centralization Measures** | | | |
| Degree | 4.75 | 9.55 | 5.0 |
| Closeness | 61.82 | 48.89 | 63.53 |
| Betweenness | 34.74 | 36.52 | 40.41 |

## 4.2. Top Growing Collaborative Countries

In order to identify the fastest growing collaborating countries in "steel structures" research, we have compared average collaborations per year for the countries that has at least one collaboration during the two periods. We defined growth rate as the division of the average number of collaborations per year during the two periods. Table 12 shows the 26 countries that have collaborations during both second and third periods in decreasing order of growth rate followed by the average number of collaborations per year for the second period (AV2) and the first period (AV1). It follows that, Portugal, France, Czech Republic, South Korea and Bangladesh have the highest collaborative growth rates.

Table 12. Top growing collaborating countries

|   | Country | Growth | AVC2 | AVC1 |   | Country | Growth | AVC2 | AVC1 |
|---|---|---|---|---|---|---|---|---|---|
| 1 | Portugal | 61.4 | 4.1 | 0.1 | 14 | United Kingdom | 6.2 | 12.4 | 2.0 |
| 2 | France | 36.8 | 7.4 | 0.2 | 15 | Netherlands | 5.8 | 3.1 | 0.5 |
| 3 | Czech Republic | 30.0 | 4.0 | 0.1 | 16 | Malaysia | 5.5 | 0.7 | 0.1 |
| 4 | South Korea | 19.1 | 6.4 | 0.3 | 17 | India | 4.5 | 0.9 | 0.2 |
| 5 | Bangladesh | 15.0 | 1.0 | 0.1 | 18 | Singapore | 4.5 | 3.3 | 0.7 |
| 6 | Spain | 14.3 | 1.9 | 0.1 | 19 | Japan | 4.4 | 7.4 | 1.7 |
| 7 | China | 11.7 | 24.1 | 2.1 | 20 | Iran | 3.8 | 1.0 | 0.3 |
| 8 | Egypt | 11.6 | 1.5 | 0.1 | 21 | Belgium | 2.9 | 2.9 | 1.0 |
| 9 | Australia | 10.1 | 12.1 | 1.2 | 22 | Turkey | 2.2 | 0.7 | 0.3 |
| 10 | Canada | 8.4 | 9.0 | 1.1 | 23 | Thailand | 2.0 | 0.3 | 0.1 |
| 11 | United States | 8.0 | 31.8 | 4.0 | 24 | Bahrain | 1.4 | 0.1 | 0.1 |
| 12 | Germany | 7.7 | 6.2 | 0.8 | 25 | Greece | 1.0 | 0.8 | 0.8 |
| 13 | Italy | 7.4 | 3.5 | 0.5 | 26 | Luxembourg | 0.2 | 0.3 | 1.6 |

## 5. Discussion and Conclusion

Using publication data and extracting co-authorship relations, we have presented an overview of collaboration efforts and collaborative networks in the "steel structures" research area. The collaboration networks of scientists in "steel structures" have been analyzed by using author affiliations from publications having 'steel structures' in their 'title' or 'keywords' or 'abstract' since 1970 to 2009 as extracted from Scopus. The publication dataset we have extracted does not support to represent the complete world production of research on "steel structures" (due to the possibility of significant biases: evolving list of relevant journals; publications in other journals and languages; and etc.). Hence, the database does not pretend to represent the complete field. Nevertheless, the paper presents a fairly accurate network of collaborations in the field of "steel structures", and illustrates the use of network theory indicators to analyse this field.

The frequency of publications shows that a few numbers of authors publish a large number of papers while the majority of authors publish a small number of papers. Thus, the distribution follows a scale-free power-law distribution. We have provided a methodological approach by applying known network and statistical measures to explore scientific collaboration networks at micro- (researcher), meso- (institute), and macro- (country) levels over the last 40 years. We find a number of interesting properties of these networks. By the growth of network size, density, connectedness, centralization measures and clustering coefficient are decreasing to the lowest values at micro-level. On the other hand, number of components and the average distance among nodes in the networks are increasing by the growth of network size.

Although the process of scientists introducing their collaborators to one another is an important one in the development of scientific communities but this had not applied in practice among weakly-clustered researchers in "steel structures" field. The low clustering coefficients values for all levels indicates that two randomly chosen authors (or institutes or countries) from the community are less likely to have collaborated if they have a third common collaborator.

Exploring average distance between countries indicate that a randomly chosen country can reach to another country just by two steps while average distance between institutes is five and authors is eight. Therefore, any two authors in "steel structures" field are connected through about eight intermediate authors. While the authors' average distance for "steel structures" field is higher than some of other fields but still it reflect a "small-world" phenomena.

Network closeness centralization is the only measure which is almost significantly high in all levels of analysis though this value is much higher in marco-level and less in meso-level and the lowest in micro-level. This reflects that the scientific collaborations network structures are centralized around few authors with high c*loseness* centrality. Networks are not centralized considering degree and betweenness centrality but they follow the same trend by having highest value in macro-level and go down for subsequent levels (i.e., meso- and micro-levels).

Investigating the network statistics during the growth of international collaboration networks over three periods of time (e.g., 1970-1982, 1983-1997, and 1998-2009) indicates increasing trend for international network's density, connectedness, and clustering coefficients. While the closeness and betweenness centralization measure are increasing too but degree centralization was decreasing. It shows that by the growth of network size, the variance between nodes' (countries) degree centrality (number of directly connected nodes) decrease while the variance between nodes' closeness and betweenness centrality increase (a few of them having high measure but a large number of them, on the edge of the networks, with low measures. Interestingly, although countries' average distance increase during second period but again decrease during third period.

## Acknowledgment

The authors appreciate the anonymous reviewers for their positive and useful comments on the early drafts of this paper.

## References

Abbasi, A. and J. Altmann (2010). A Social Network System for Analyzing Publication Activities of Researchers. Symposium on Collective Intelligence (COLLIN 2010), Advances in Intelligent and Soft Computing, Hagen, Germany, Springer.
Abbasi, A. and J. Altmann (2011). On the Correlation between Research Performance and Social Network Analysis Measures Applied to Research Collaboration Networks. Hawaii International Conference on System Sciences, Proceedings of the 44th Annual., Waikoloa, HI, IEEE.
Abbasi, A., J. Altmann and L. Hossain (in press). "Identifying the Effects of Co-Authorship Networks on the Performance of Scholars: A Correlation and Regression Analysis of Performance Measures and Social Network Analysis Measures." Journal of Informetrics.

Abbasi, A., J. Altmann and J. Hwang (2010). "Evaluating scholars based on their academic collaboration activities: two indices, the RC-index and the CC-index, for quantifying collaboration activities of researchers and scientific communities." Scientometrics **83**(1): 1-13.

Acedo, F. J., C. Barroso, C. Casanueva and J. L. Galán (2006). "Co Authorship in Management and Organizational Studies: An Empirical and Network Analysis*." Journal of Management Studies **43**(5): 957-983.

Barabási, A. L., H. Jeong, Z. Néda, E. Ravasz, A. Schubert and T. Vicsek (2002). "Evolution of the social network of scientific collaborations." Physica A: Statistical Mechanics and its Applications **311**(3-4): 590-614.

Bavelas, A. (1950). "Communication patterns in task-oriented groups." Journal of the Acoustical Society of America **22**: 725-730.

Borgatti, S. (1995). "Centrality and AIDS." Connections **18**(1): 112-114.

Borgatti, S., M. Everett and L. Freeman (2002). "Ucinet for windows: Software for social network analysis (version 6)." Harvard, MA: Analytic Technologies.

Freeman, L. C. (1979). "Centrality in social networks conceptual clarification." Social Networks **1**(3): 215-239.

Freeman, L. C. (1980). "The gatekeeper, pair-dependency and structural centrality." Quality and Quantity **14**(4): 585-592.

Grossman, J. W. (2002). "The evolution of the mathematical research collaboration graph." Congressus Numerantium: 201-212.

Jiang, Y. (2008). "Locating active actors in the scientific collaboration communities based on interaction topology analyses." Scientometrics **74**(3): 471-482.

Katz, J. and B. Martin (1997). "What is research collaboration?" Research policy **26**(1): 1-18.

Leclerc, M. and J. Gagné (1994). "International scientific cooperation: The continentalization of science." Scientometrics **31**(3): 261-292.

Luukkonen, T., O. Persson and G. Sivertsen (1992). "Understanding patterns of international scientific collaboration." Science, Technology & Human Values **17**(1): 101.

Luukkonen, T., R. Tijssen, O. Persson and G. Sivertsen (1993). "The measurement of international scientific collaboration." Scientometrics **28**(1): 15-36.

Melin, G. (2000). "Pragmatism and self-organization: Research collaboration on the individual level." Research policy **29**(1): 31-40.

Melin, G. and O. Persson (1996). "Studying research collaboration using co-authorships." Scientometrics **36**(3): 363-377.

Moody, J. (2004). "The structure of a social science collaboration network: Disciplinary cohesion from 1963 to 1999." American Sociological Review **69**(2): 213.

Newman, M. E. J. (2001). "Scientific collaboration networks. I. Network construction and fundamental results." Physical review E **64**(1): 16131.

Newman, M. E. J. (2001). "Scientific collaboration networks. II. Shortest paths, weighted networks, and centrality." Physical review E **64**(1): 16132.

Newman, M. E. J. (2001). "The structure of scientific collaboration networks." Proceedings of the National Academy of Sciences of the United States of America **98**(2): 404.

Newman, M. E. J. (2004). "Coauthorship networks and patterns of scientific collaboration." Proceedings of the National Academy of Sciences of the United States of America **101**(Suppl 1): 5200.

Owen-Smith, J., M. Riccaboni, F. Pammolli and W. Powell (2002). "A comparison of US and European university-industry relations in the life sciences." Management Science **48**(1): 24-43.

Sabidussi, G. (1966). "The centrality index of a graph." Psychometrika **31**(4): 581-603.

Scott, J. (1991). Social network analysis: a handbook., Sage.

Sonnenwald, D. (2007). "Scientific collaboration: a synthesis of challenges and strategies." Annual Review of Information Science and Technology **41**: 643-681.

Suresh, V., N. Raghupathy, B. Shekar and C. E. Madhavan (2007). "Discovering mentorship information from author collaboration networks." Discovery Science **4755**: 197-208.

US Office of Science & Technology Policy (2000). Examples of international scientific collaboration and the benefits to society, Retrieved June 14, 2005 from http://clinton4.nara.gov/WH/EOP/OSTP/html/00426_7.html.

Wagner, C. S. and L. Leydesdorff (2005). "Network structure, self-organization, and the growth of international collaboration in science." Research Policy **34**(10): 1608-1618.

Watts, D. and S. Strogatz (1998). "Collective dynamics of 'small-world' networks." Nature **393**(6684): 440-442.